\documentclass[preprint]{aastex}

\newcommand{\sbs}{SBS~0335--052}

\begin{document}
\title{Abundances in the H {\sc i} envelope of the extremely low-metallicity
 Blue Compact Dwarf Galaxy SBS~0335--052 from {\it Far Ultraviolet Spectroscopic Explorer} observations}
\author{Trinh X. Thuan}
\affil{Astronomy Department, University of Virginia,
P.O. Box 3818, University Station, Charlottesville, VA, 22903}
\email{txt@virginia.edu}
\author{Alain Lecavelier des Etangs}
\affil{Institut d'Astrophysique de Paris, CNRS, 98 bis Boulevard Arago,
75014 Paris, France}
\email{lecaveli@iap.fr}
\and
\author{Yuri I. Izotov}
\affil{Main Astronomical Observatory, National Academy of Sciences,
27 Zabolotnoho str., 03680 Kyiv, Ukraine}
\email{izotov@mao.kiev.ua}

\altaffiltext{}{Based on observations obtained with the NASA-CNES-CSA {\sl Far Ultraviolet Spectroscopic Explorer}. {\sl FUSE} is operated for NASA by the Johns Hopkins University under NASA contract NAS 5-32985.}

\begin{abstract}
We present {\sl FUSE} spectroscopy of SBS\,0335--052, 
the second most metal-deficient blue compact dwarf (BCD) galaxy known 
(log O/H = --4.70).
In addition to the H\,{\sc i} Lyman series, we detect C\,{\sc ii}, N\,{\sc i},
N\,{\sc ii}, O\,{\sc i}, Si\,{\sc ii}, Ar\,{\sc i} and Fe\,{\sc ii} 
absorption lines, mainly arising from 
the extended H\,{\sc i} envelope in which \sbs\ is embedded.
 No H$_2$ absorption lines are seen. The absence of diffuse H$_2$
implies that the
warm H$_2$ detected through infrared emission must be very clumpy and 
associated with the star-forming regions.
The clumps should be denser than $\sim$ 1000 cm$^{-3}$ and hotter than 
$\sim$ 1000 K and account for $\geq$ 5\% of the total H\,{\sc i} mass. 
Although \sbs\ is a probable
young galaxy, its neutral gas is not pristine. The metallicity of its 
neutral gas is similar to that of its ionized gas and is 
equal log O/H $\sim$ --5.
This metallicity is comparable to those found in the 
 H\,{\sc i} envelopes of 
four other BCDs with ionized gas metallicities
spanning the wide range from log O/H = --4.8 to log O/H = --3.8, and in 
Ly$\alpha$ absorbers, fueling the speculation that there may have been 
previous enrichment of the primordial neutral gas to a common metallicity 
level of log O/H $\sim$ --5, possibly by Population III stars.

\end{abstract}


\section{Introduction}

The Blue Compact Dwarf (BCD) galaxy SBS 0335--052 \citep{I90}, with an ionized gas
oxygen abundance of
12 + log O/H = 7.30 \citep{M92,I97,I99} 
[$Z_\odot$/23, if the recent lower determination 
of the Oxygen abundance of the Sun 12 + log O/H = 8.66 by \citet{A04} 
is adopted], is
one of the most metal-deficient star-forming galaxies known, just after
its dwarf irregular companion  SBS 0335--052W \citep[12 + log O/H = 7.22 
or $Z_\odot$/28,][] {L99} and
the BCD I~Zw~18 \citep[12 + log O/H = 7.17 or $Z_\odot$/31,][]{I99}.
Because of its extremely low metallicity and its
high neutral hydrogen content \citep{P01}, \sbs\ is an excellent
candidate for being a young galaxy, just like I Zw 18.
The latter has been shown by \citet{IT04} to be a bona fide young
galaxy. Resolving I Zw 18 into stars with deep {\sl Hubble Space Telescope}
({\sl HST})/Advanced Camera for Survey images and
constructing its color-magnitude diagram (CMD),
\citet{IT04} have shown that
the galaxy does not contain red giant stars and that the most evolved stars
in it are not older than 500 Myr.
Because of its larger distance 
\citep[54.3 Mpc,][ hereafter TIL97, instead of 15 Mpc for I Zw 18]{TIL97}, 
SBS 0335--052 cannot be resolved into
stars by {\sl HST}, so its age cannot be determined directly by CMD analysis.
However, using optical
colors of the low surface brightness component together with
evolutionary synthesis models,  TIL97 and
\citet{P98} have constrained the age of the underlying stellar
population to be less than $\sim$ 100 Myr. Near-infrared (NIR) colors after
correction for gaseous emission are consistent with a stellar population
not older than 4 Myr, with a possible contribution to the NIR light
from an evolved stellar population not exceeding $\sim$ 15\%
\citep{V00}. \citet{OK01} have derived from surface photometry 
considerably larger ages for \sbs. However these large ages 
may be the consequence of their not subtracting out the 
large ionized gas contamination from the surface brightness profiles 
used. Recently, \citet{P04} have also found a young age 
for \sbs. Depending on the adopted star formation
history of the BCD, they derive an age between 100 Myr and 400 Myr.
 In any case, \sbs\ constitutes an excellent nearby laboratory for
studying massive star formation and its interaction with the
ambient interstellar medium (ISM) in a very metal-deficient environment.
Most of the star formation in \sbs\ occurs in six super-star clusters,
roughly aligned in the southeast-northwest direction, with
ages $\leq$ 25 Myr, within a region of $\sim$ 2$\arcsec$ or 520 pc in size
(TIL97).

VLA observations have shown the presence of a
massive (2.2$\times$10$^9$ $M_\odot$) and extended (66 kpc$\times$22 kpc)
neutral hydrogen gas envelope around SBS 0335--052 \citep{P01}.
This H\,{\sc i} envelope absorbs all the Ly$\alpha$
photons emitted by the young massive stars.
\citet{TI97} have obtained a {\sl HST}/Goddard High Resolution Spectrograph (GHRS) 
spectrum of
\sbs\ which reveals a broad damped
Ly$\alpha$ absorption with a high H\,{\sc i} column density
$N$(H\,{\sc i}) = (7.0$\pm$0.5)$\times$10$^{21}$ cm$^{-2}$, the largest known in
a BCD. The {\sl HST}/GHRS spectrum also shows absorption lines of several
heavy elements such as O {\sc i}, Si {\sc ii} and S {\sc ii}. Assuming
that these absorption lines are not saturated, \citet{TI97} found
that the derived heavy element abundance in the neutral gas to be
considerably lower than that of the ionized gas.
Furthermore, \citet{V00} have detected with infrared spectrophotometry
molecular hydrogen in emission in \sbs.
The presence of a neutral gas envelope around SBS 0335--052 and the
detection of molecular hydrogen in emission make this galaxy an ideal
target for {\sl FUSE} spectroscopy. Using \sbs\ as a source of UV light
shining through the H {\sc i} envelope,
a {\sl FUSE} absorption spectrum will allow us to determine
independently the heavy element abundances in the neutral gas of 
the galaxy and check for the presence of H$_2$ in it.

We describe the {\sl FUSE} observations in Section 2.  In Section 3, we set
upper limits on the amount of diffuse H$_2$. In Section 4, we 
show that the warm H$_2$ detected through infrared emission must be very 
clumpy. We derive the column densities of the interstellar
ionic species in the H {\sc i} gas,
and compare the heavy element abundances in the neutral
and ionized gas in Section 5. We summarize our findings in Section 6.

\section{Observations and data analysis}

\sbs\ has been observed with {\sl FUSE} \citep{M00} through the
30\arcsec$\times$30\arcsec\ LWRS large entrance aperture on 2001, September 26 for a
total integration time of 24.4~ksec. The data have been processed with
version 1.8.7 of the {\tt CALFUSE} pipeline. The eight separate exposures
have been aligned and coadded. This results in a set of four independent
spectra, two spectra for the two long wavelength LiF channels covering the
wavelength range $\sim$ 1000-1187 \AA, and two spectra for the two short
wavelength SiC channels, covering the wavelength range  $\sim$ 900-1100
\AA. After co-addition, the S/N ratio per resolution element and per
channel is estimated to be about~2 below 1000~\AA\ and about~4 above
1000~\AA.

To derive the spectrum, special attention was paid to the background
residual which is not negligible for this faint target. The background has
been estimated both on spectra obtained off-target through other apertures
and at the bottom of wide saturated lines. The background level is found
to be between 0 and $3\times 10^{-15}$ erg~cm$^{-2}$~s$^{-1}$~\AA$^{-1}$.
The line spread function has been estimated with the unresolved Galactic
H$_2$ absorption lines. The resolution is found to be $R=\Delta
\lambda/\lambda \approx 12000$.

\begin{figure}
\epsscale{0.8}
\plotone{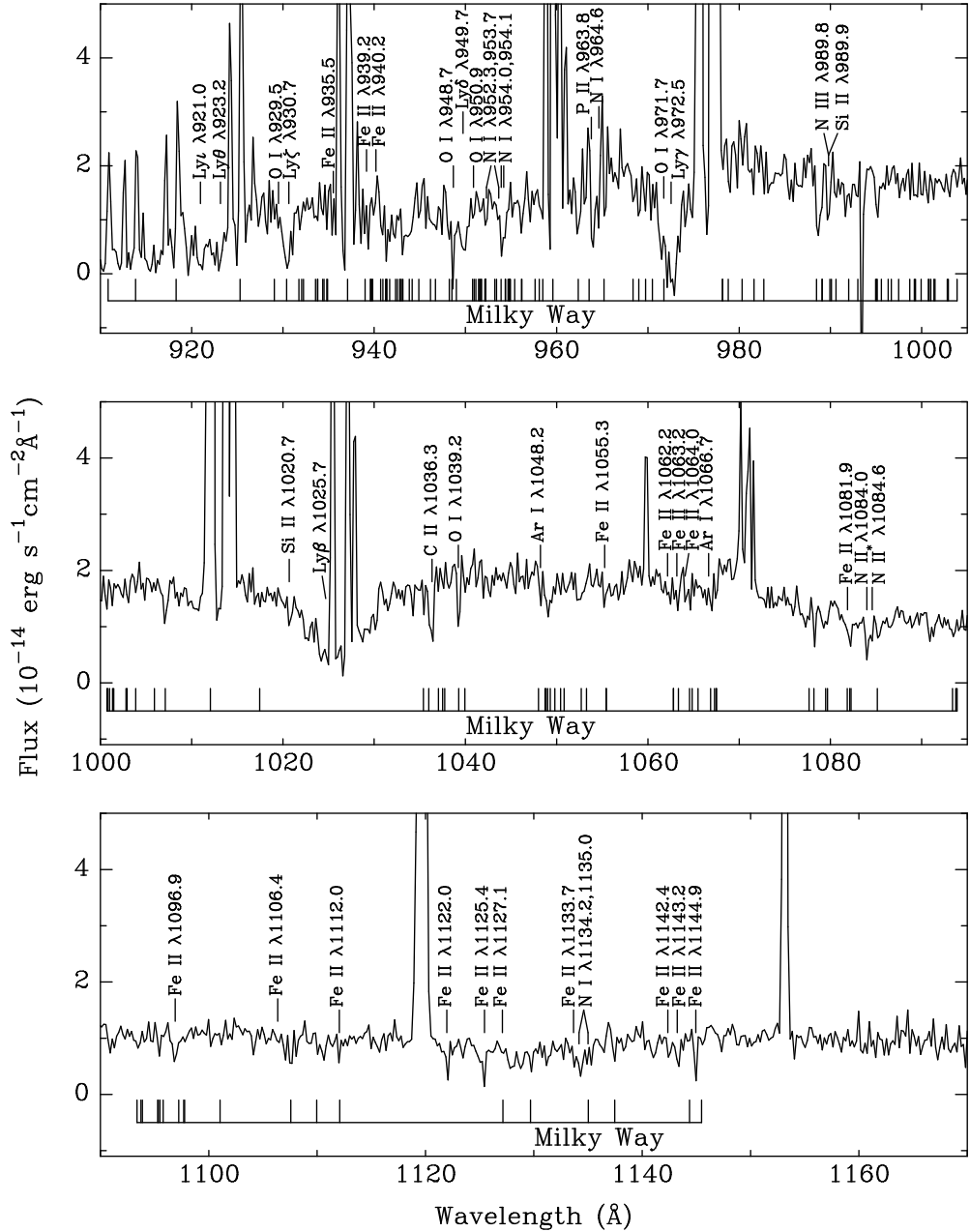} 
\figcaption{
{\sl FUSE} spectrum in the rest frame of \sbs\ (shifted
by 4050~km~s$^{-1}$).
Here the data have been rebinned to a resolution of 0.2~\AA .
Prominent interstellar absorption lines are indicated.
The lines arising in \sbs\ are marked on top and those in
the Milky Way at bottom.
In addition to the H\,{\sc i} Lyman series in \sbs,
there are also strong interstellar absorption lines from
C {\sc ii}, N {\sc i}, N {\sc ii}, O {\sc i},
Si {\sc ii} and Fe {\sc ii}. \label{fig1}}
\end{figure}

The resulting spectrum of \sbs\ shifted to its rest-frame is shown in
Fig. \ref{fig1}. It exhibits two absorption-line systems at two different radial
velocities. The first system at nearly zero radial velocity is attributed
to the interstellar clouds in the Milky Way. The second system at
$\sim$4050~km~s$^{-1}$ is attributed to the ISM in \sbs, as the H\,{\sc i}
velocity of the BCD is 4057$\pm$5~km~s$^{-1}$ \citep{P01}, in
good agreement with the optical velocity obtained by \citet{I97}.
Absorption lines from the atomic and ionic species N\,{\sc i}, P\,{\sc
ii}, Fe\,{\sc ii}, and Ni\,{\sc ii} are observed in the Milky Way, while
for \sbs, in addition to the H {\sc i} Lyman series, absorption lines from
the atoms and ions C\,{\sc ii}, N\,{\sc i}, N\,{\sc ii}, O\,{\sc i},
Si\,{\sc ii}, Ar\,{\sc i}, and Fe\,{\sc ii} are seen. In Fig. \ref{fig1}, the
lines arising from \sbs\ are labeled while those belonging to the Milky
Way are indicated by tickmarks below the {\sl FUSE} spectrum.
For the characteristics of the electronic transitions of H$_2$, 
we have used the wavelengths and
oscillator strengths tabulated by 
\citet{A93a,A93b} respectively for the Lyman and the Werner systems, 
and the inverses of the total
radiative lifetimes tabulated by \citet{A00}. For atomic lines,
we have used the data for resonance absorption lines tabulated by \citet{M03}.

\section{Upper limits on the diffuse H$_2$ content of SBS 0335--052}

Molecular hydrogen is known to be present in \sbs\ as \citet{V00}
have detected several H$_2$ emission lines in their low-resolution
near-infrared spectrum of the star-forming region. These lines observed in
the $K$ band are generally consistent, within the errors, with both
thermal and fluorescent excitation by the strong UV field. It is thus of
great interest to check whether there are H$_2$ absorption lines in the
{\sl FUSE} spectrum of \sbs. We found no H$_2$ line at the radial velocity
of the BCD (Fig. \ref{fig1}). We obtained the following 3 $\sigma$ upper limits for
the H$_2$ column density: 1.2$\times$10$^{17}$ cm$^{-2}$,
1.4$\times$10$^{17}$ cm$^{-2}$, 7.9$\times$10$^{16}$ cm$^{-2}$,
3.4$\times$10$^{17}$ cm$^{-2}$ and 1.5$\times$10$^{17}$ cm$^{-2}$
respectively for the J= 0, 1, 2, 3 and 4 levels. To set an upper limit on
the total H$_2$ column density, we follow the procedure of \citet{A03}.
We add the two upper limits for the J=0 and J=1 levels. For the
higher J levels, we assume a temperature ($T$ = 500 K and $T$ = 1000 K), which
determines the level populations, and then normalize the level populations
so as to be consistent with the measured upper limits. We obtain upper
limits for the total H$_2$ column density of 3.7$\times$10$^{17}$
cm$^{-2}$ for $T$ = 500 K and 5.5$\times$10$^{17}$ cm$^{-2}$ for $T$ = 1000 K.
These upper limits for H$_2$ are more than two orders of magnitude larger
than those established for the BCDs I Zw 18 \citep{A03,L04} and Mrk 59 \citep{T02}
because of
the larger distance of \sbs, its fainter apparent magnitude and its lower
signal-to-noise ratio {\sl FUSE} spectrum. They correspond to column densities 
above which the damping wings start to 
broaden appreciably the H$_2$ absorption lines. 
With a H {\sc i} column density of
7.0$\times$10$^{21}$ cm$^{-2}$ as determined by the {\sl HST}/GHRS damped Ly$\alpha$
profile \citep{TI97}, this corresponds to a fraction of H$_2$ molecules  
$f({\rm H}_2)$ = 1.1$\times$10$^{-4}$ ($T$ = 500 K) and $f({\rm H}_2)$ = 
1.6$\times$10$^{-4}$ ($T$ = 1000 K) where the fraction is 
defined as $f$(H$_2$) =
2$n$(H$_2$)/[2$n$(H$_2$)+$n$(H {\sc i})],  and 
where $n$(H$_2$) and $n$(H {\sc i}) are 
the number densities of H$_2$ and H {\sc i}.

Are the H$_2$ upper limits established by {\sl FUSE} 
consistent with the expected fraction of H$_2$
molecules in \sbs? This fraction can be estimated in the following 
manner. As discussed by \citet{Vi00}, at equilibrium, when the
formation of H$_2$ on dust equals its destruction rate by UV photons,
then:
\begin{equation}
f({\rm H}_2)=1.6\times10^{-34}\left(\frac{Z}{Z_\odot}\right)
\left(\frac{F_{R_0}}{{\rm erg\ s^{-1}\ cm^{-2}\ \AA^{-1}}}\right)^{-1}
\left(\frac{R_0}{\rm kpc}\right)^{-1}\left[\frac{N({\rm HI})}{\rm cm^{-2}}\right].
\label{eq1}
\end{equation}
Here $F_{R_0}$ is the flux at 1000\AA\ of UV H$_2$ dissociating radiation
at radius $R_0$ from the central ionizing clusters in SBS 0335--052, $Z$ is
the metallicity equal to $Z_\odot$/23, and $N$(H {\sc i}) is the H {\sc i} column
density. From the H {\sc i} map of \citet{P01}, we estimate
$R_0$ = 10\arcsec, or 2.5 kpc for a H {\sc i} column density level of
7.2$\times$10$^{20}$ cm$^{-2}$. This corresponds to 
a H {\sc i} number density $n$(H {\sc i}) $\sim$
0.1 cm$^{-3}$. Since the density of the H {\sc i} is increasing inwards,
we adopt for our calculations a mean value $n$(H {\sc i}) = 1 cm$^{-3}$.

The observed UV flux at 1000\AA\ is 1.7$\times$10$^{-14}$
erg s$^{-1}$ cm$^{-2}$ \AA$^{-1}$ \citep{TI97}.
Correcting for interstellar extinction
using $A$(1000\AA) = 5.742$A_V$ \citep{C89} and
$A_V$ = 0.155 mag \citep{S98}, we obtain $F_{cor}$ =
3.8$\times$10$^{-14}$ erg s$^{-1}$ cm$^{-2}$ \AA$^{-1}$. Then
\begin{equation}
F_{R_0}=F_{cor}\left(\frac{D}{R_0}\right)^2 =
1.7\times10^{-5}\ {\rm erg\ s^{-1}\ cm^{-2}\ \AA^{-1}}, \label{eq2}
\end{equation}
where $D$ = 54.3 Mpc is the distance of SBS 0335--052. Finally, from Eq. \ref{eq1}
we derive $f$(${\rm H}_2$) = 1.1$\times$10$^{-10}$, more than six orders of
magnitude below and fully consistent with the upper limits 
 derived from the {\sl FUSE} spectrum.

\section{A clumpy interstellar medium}

Is such a low fraction of H$_2$ consistent with the detection of
near-infrared (NIR) H$_2$ emission lines in the star-forming region of \sbs\
by \citet{V00}? We note that such a question has also arisen in the context of 
a circumstellar disk observed by \citet{Lecav2001} where H$_2$ is 
detected in the infrared, but where no UV absorption lines are seen. 
To address the question for \sbs, we have
carried out calculations specifically for the H$_2$(1,0)S(0) line
with the wavelength of 2.223 $\mu$m. The results for the other detected 
H$_2$ lines should not be too different. \citet{V00}
have obtained a flux of 1.4$\times$10$^{-16}$ erg s$^{-1}$ cm$^{-2}$
for the 2.223 $\mu$m line in \sbs.
Assuming that the NIR emission is the result of fluorescent excitation,
the luminosity of the H$_2$(1,0)S(0)
line from the region with radius $R_0$ is
\begin{equation}
L_{tot}=n({\rm HI})f({\rm H_2})Qh{\nu}V, \label{eq3}
\end{equation}
where $h\nu$ = 9$\times$10$^{-13}$ erg is the energy of the photon with
wavelength 2.223$\mu$m, and
$V$=4$\pi$$R_0^3$/3 = 1.92$\times$10$^{66}$ cm$^{-3}$
is the volume of the region with radius $R_0$. 
Following \citet{BD76} and using the UV flux $F_{R_0}$ given by 
Eq. \ref{eq2}, we calculate a cascade entry rate $Q$ =
1.55$\times$10$^{-11}$ cm$^{-3}$ s$^{-1}$.
We obtain from Eq. \ref{eq3} the
luminosity of the 2.223$\mu$m line to be
$L_{tot}$=1.82$\times$10$^{33}$ erg s$^{-1}$.
The spectrum of \citet{V00} was obtained through a 
1\arcsec$\times$1\farcs5
aperture, which corresponds to a volume
$V_{Vanzi}$ = 1.38$\times$10$^{64}$ cm$^{-3}$. 
Thus, $V_{Vanzi}$/$V$ = 7.2$\times$10$^{-3}$ 
and we would predict $L_{predicted}$=1.31$\times$10$^{31}$ erg s$^{-1}$,
corresponding to a flux at Earth
of 2.4$\times$10$^{-23}$ erg s$^{-1}$cm$^{-2}$. This is
$\sim$ 10$^{7}$ times lower than the observed flux.

The large flux discrepancy comes very probably from our assumption that the
interstellar medium of \sbs\ is uniform. In all likelihood, the
NIR H$_2$ emission comes,
not from an uniform low-density neutral medium, but from dense clumps.
For such dense clumps, self-shielding
of electronic transitions, which becomes important starting at $N$(H$_2$)
$\sim$ 10$^{14}$ cm$^{-2}$ \citep{BvD87}, further increases
$f$(H$_2$). Although the UV pumping decreases,
thermal excitation of the H$_2$ vibrational states becomes important 
since it scales as $n$(H)$\times$$n$(H$_2$). Following \citet{BvD87},
we find that we can reproduce well the observed flux of the H$_2$(1,0)S(0) 
2.223 $\mu$m line if it comes from an ensemble of clumps with 
a 
total mass $\sim$ 10$^8$ $M_\odot$, or about 5\% of the total H {\sc i} mass 
in \sbs\ of 2.1$\times$10$^{9}$ $M_\odot$ \citep{P01}, and if 
each clump has a  
number density $n$(H {\sc i}) = 5$\times$10$^3$ cm$^{-3}$ and a 
temperature $T$ = 10$^3$ K.  For such a clump to be stable against
gravitational collapse, its mass has to be less than its Jeans mass
which is $\sim$ 5$\times$10$^{4}$ $M_\odot$. This means that there 
should be at least $\sim$ 2000 of these clumps.
The fraction
of H$_2$ molecules in such clumps is $f$(H$_2$) = 5$\times$10$^3$ $\times$
6.8 $\times$ 10$^{-11}$ = 3.4 $\times$ 10$^{-7}$. The total number
of H$_2$ molecules in such clumps is $N_{tot}$ = $f$(H$_2$)$M$/$m_{\rm H}$ =
4.1$\times$10$^{58}$, where $m_{\rm H}$ = 1.673$\times$10$^{-24}$ g is 
the mass
of the hydrogen atom. Then the luminosity emitted in the H$_2$(1,0)S(0) line
is $L$=$N_{tot}$$n$(H {\sc i})$q$$h\nu$=4.5$\times$10$^{37}$ erg s$^{-1}$,
where $q$ $\sim$ 10$^{-10}$exp(--6000/$T$) cm$^{3}$s$^{-1}$ is the rate of
collisional excitation.
The flux at Earth is then 1.4$\times$10$^{-16}$ erg s$^{-1}$ cm$^{-2}$ in
very good agreement with the observed flux for the  H$_2$(1,0)S(0) line
at 2.223 $\mu$m.
 Because of the exponential temperature dependence of the rate 
of collisional excitation, the above
estimate is very sensitive to the adopted temperature for the gaseous clumps.
This temperature cannot be lower than $\sim$ 1000 K,
since for lower temperatures, 
 the total mass of the clumps needed to account for the observed 
H$_2$(1,0)S(0) 2.223 $\mu$m flux 
would exceed the H {\sc i} mass of \sbs.

Our {\sl FUSE} observations are not sensitive to such a clumpy H$_2$ distribution.
They can only probe diffuse H$_2$ along the line of sights to the several
thousands of UV-bright massive stars in \sbs.
As in the cases of the BCDs I Zw 18 \citep{A03,L04}
and Mrk 59 \citep{T02}, the absence of diffuse
H$_2$ in \sbs\ can be explained by the combined effects
 of a low H {\sc i} density,
a scarcity of dust grains on which H$_2$ molecules can form, and a large
UV flux which destroys molecules.

\section{Heavy element abundances}

\subsection{Column densities}

We now derive the column densities of the heavy elements in the neutral H {\sc i}
envelope of \sbs\ by fitting the profiles of the metal absorption lines.
Column densities have been calculated by profile fitting using
the {\tt Owens} procedure developed by Martin Lemoine and the {\sl FUSE}
French team. This code returns the most likely values of many free
parameters like the Doppler widths and column densities through a $\chi^2$
minimization of the difference between the observed and computed profiles.
The latest version of this code is particularly suited to the
characteristics of {\sl FUSE} spectra. For example, it allows for a
variation of the background level, for an adjustable line spread function
as a function of the wavelength domain, and for shifts in wavelength
scale. These are taken as free parameters which depend on the wavelength
region and are determined by a $\chi^2$ minimization.

The derived column densities with their error 
bars are given in Table \ref{Tab1}. We will be quoting 1 $\sigma$ error bars 
throughout the paper.
These have been estimated as usual by considering the $\Delta
\chi^2$ increase of the $\chi^2$ of the fit 
\citep[see][ for a full discussion of the fitting method and
error estimation with the {\tt Owens} code]{H02}. These error bars include the
uncertainties in the continuum, the intrinsic line widths ($b$), the
background residual and the instrumental line spread function. It is
important to remark that the use of different lines with different oscillator
strengths of the same species allows us to constrain reasonably well 
all these quantities.
In particular, the line width is well constrained by strongly saturated 
lines. On the other hand, the instrumental line spread function is mainly
constrained by the 
very narrow lines from H$_2$ in the
Milky Way. The final results are obtained from a simultaneous
self-consistent fit to all the data. In this final fit, the continuum, the
background residual, the instrumental line spread function, and the physical
parameters of the absorption lines are free parameters. They are estimated by
determining the best $\chi^2$ in the parameters' space.

For each wavelength domain, we use simultaneously the data from two
different channels: the SiC~1 and SiC~2 channels for wavelengths between
940 and 990~\AA, and the LiF~1 and LiF~2 channels for wavelengths above $\sim
990$~\AA\ \citep[see][ for a description of the {\sl FUSE} channels]{S00}. 
Because of the low signal-to-noise ratio of the \sbs\ {\sl FUSE} 
spectrum, only lines that are more or less saturated are clearly
detected. This explains the relatively
large error bars on the estimated column densities. However, for most
species detected in SBS\,0335--052, we have for each species 
several absorption lines from
different transitions with different oscillator strengths. This allows us to
obtain reliable column densities despite the saturation of the 
lines. For example, several O\,{\sc i} lines are 
detected. The
O\,{\sc i} column density is mainly constrained by the profiles of four
strong transitions with rest wavelengths $\lambda_0=929.5$\,\AA,
950.9\,\AA, 988.8\,\AA, and 1039.2\,\AA. For N\,{\sc i}, we detect the
well-known triplet at 1034-1035\,\AA. The ions 
N\,{\sc ii} and N\,{\sc ii}* have
transitions in the wavelength region around 1084-1085\,\AA\ which is
usually unusable because of its location 
in the hole between the two LiF detectors
\citep[see e.g.,][ in the case of I\,Zw\,18]{L04}.
However, here the large redshift of SBS\,0335--052 moves that region 
out of the hole and allows clear
detections of these transitions in the LiF\,1b and LiF\,2a channels.
Si\,{\sc ii} is detected only in the $\lambda_0=1020.7$\,\AA\ transition,
the line at $\lambda_0=989.9$\,\AA\ being blended with a strong 
N\,{\sc iii} line. However with an optical depth of $\sim$2, 
this line does not suffer from much saturation 
and allows a reliable column density determination, albeit with 
large error bars. On the other hand, C\,{\sc ii}  which is also
detected only in one transition at $\lambda_0=1036.3$\,\AA, is strongly 
saturated. This makes its column density
determination very sensitive to systematic errors and can explain its
deviation from the abundance pattern found for other species (see
Section~5.2). 

\begin{figure}
\epsscale{0.8}
\plotone{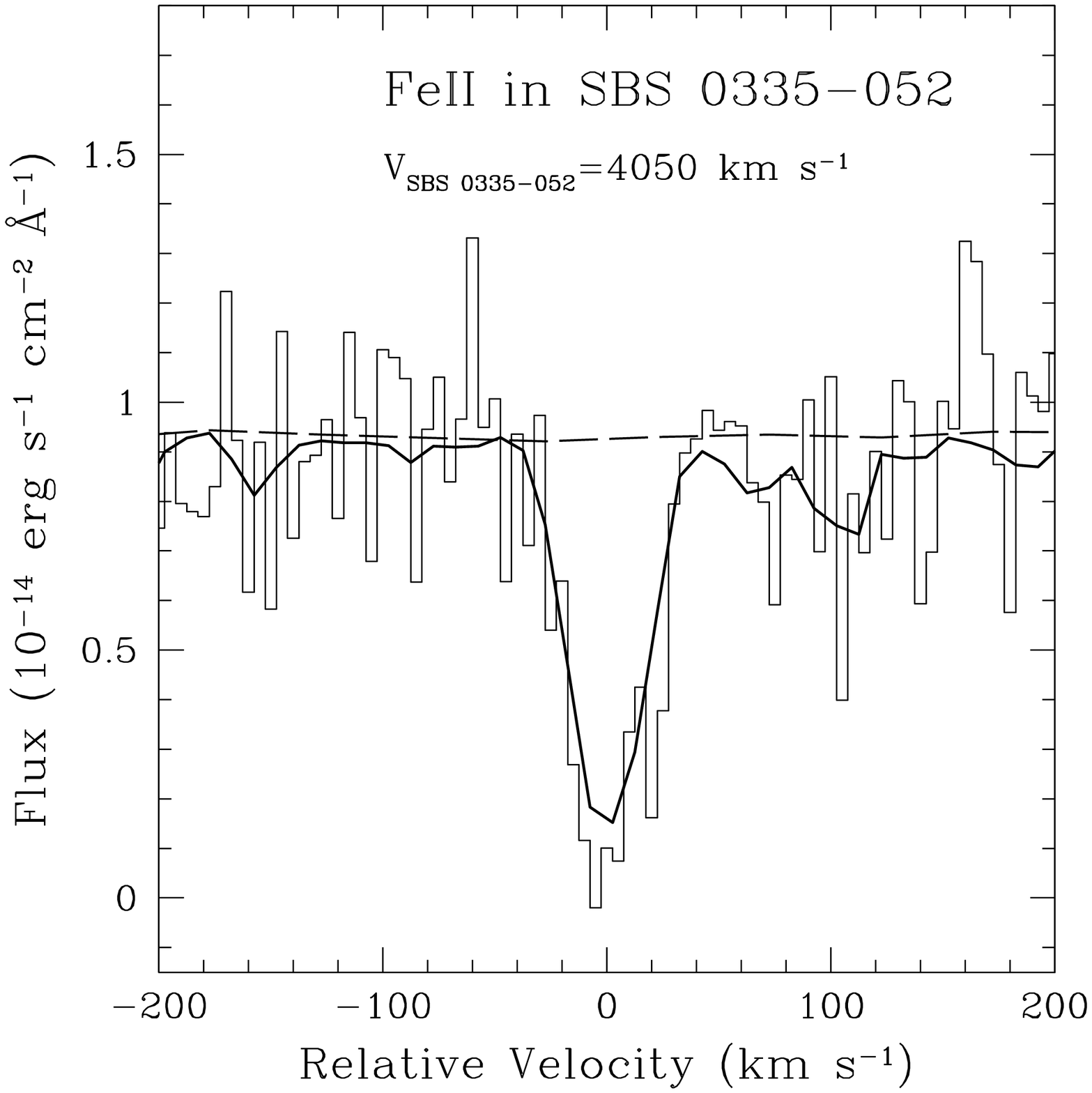} 
\figcaption{Co-added profile of the Fe\,{\sc ii} absorption 
line in \sbs . The profile and adjacent continua have
been obtained by an error-weighted co-addition of 9~different
Fe\,{\sc ii} lines at wavelengths 
$\lambda_0=1055$\,\AA, 1063\,\AA, 1064\,\AA,
1082\,\AA, 1097\,\AA, 1122\,\AA, 1125\,\AA, 1143\,\AA\ and 1145\,\AA . 
The profile fits to individual lines 
have also been co-added, and the result is shown by a thick line.
The dashed line shows the addition of the corresponding continua.
Other lines can also be seen such as a  H$_2$\,(J=1) line and 
a Fe\,{\sc ii} line from the Milky Way at $\sim$110~km~s$^{-1}$. \label{fig2}}
\end{figure}

With the exception of H\,{\sc i} for which the strong damping 
wings in the
Lyman\,$\beta$ line allow a very accurate determination of the column
density, Fe\,{\sc ii} is the species for which we have obtained the highest
precision (Table \ref{Tab1}). Fe\,{\sc ii} has indeed nine
clearly detected lines at $\lambda_0$ =1055\,\AA, 1063\,\AA, 1064\,\AA,
1082\,\AA, 1097\,\AA, 1122\,\AA, 1125\,\AA, 1143\,\AA\ and 1145\,\AA. 
Those lines have optical depths ranging from $\sim$2 to $\sim$30, 
which allows a good determination of
the Fe\,{\sc ii} column density by co-adding the nine separate 
profiles to improve the signal-to-noise ratio. 
Fig. \ref{fig2} shows the fit to the co-added continuum and Fe\,{\sc
ii} line profile. 

To better constrain the column densities, we have also used, in addition
to the {\sl FUSE} data, the low resolution {\sl HST} spectra obtained with
the GHRS spectrograph and the G160M and G140L gratings by \citet{TI97}.
These spectra show absorption lines of C\,{\sc ii} (1334.5~\AA),
O\,{\sc i} (1302.2~\AA) and Si\,{\sc ii} (1190.4, 1193.3, 1260.4 and
1304.4~\AA).

For the species which are detected in both {\sl FUSE} and {\sl HST}
spectra (O\,{\sc i}, C\,{\sc ii} and Si\,{\sc ii}), the column
densities given in Table \ref{Tab1}
are derived from a simultaneous fit of both spectra. This assumes that both
instruments cover the same spatial regions in the BCD,  although the
observations have been obtained through very different aperture sizes,
30\arcsec\ in the case of {\sl FUSE}, and 2\arcsec\ for {\sl HST}. This assumption
is reasonable as the brightest star-forming clusters in \sbs\ are confined
within a very compact region ($\la$ 2\arcsec, TIL97), and this region is fully
sampled by both {\sl FUSE} and {\sl HST} apertures. That this is indeed
the case is confirmed by the fact that the H\,{\sc i} column densities
derived independently from the {\sl FUSE} and {\sl HST} spectra are very
similar: the {\sl FUSE} column density is log $N$(H\,{\sc i}) = $21.86^{+0.08}_{-0.05}$, 
while the {\sl HST} column density is log $N$(H\,{\sc i}) = $21.85\pm 0.05$.
The heavy element column densities derived from 
the {\sl FUSE} spectra alone and from 
the {\sl FUSE+HST} spectra are consistent with each other within 
the errors. But by 
using the {\sl FUSE} and {\sl HST} spectra together, we are able to 
decrease the error bars by a factor of about 2. 
For example, the derived 
column density of C {\sc ii} using only the {\sl FUSE} spectrum is 
log $N$(C {\sc ii}) ({\sl FUSE}) = 17.68$^{+0.23}_{-1.6}$ as compared 
to log $N$(C {\sc ii}) ({\sl HST+FUSE}) = 17.59$^{+0.23}_{-0.65}$. 

\begin{figure}
\epsscale{0.8}
\plotone{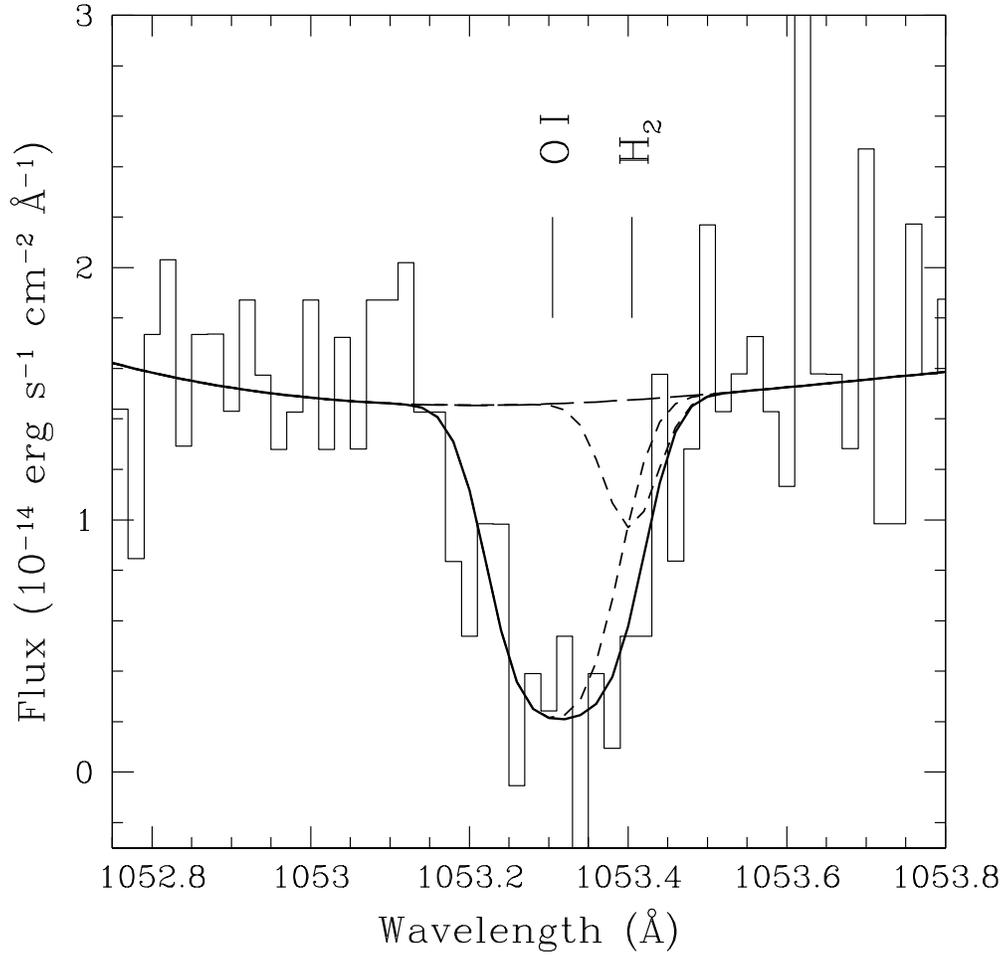} 
\figcaption{
Profile of the O {\sc i} $\lambda$1039\AA\ absorption 
line in the LiF1a channel. 
The fit is shown by small dashes while the continuum 
is shown by long dashes. The other absorption line 
at $\lambda\sim1053.4$\,\AA\ is due to Galactic H$_2$. Its fit is 
also shown by small dashes. The sum of the two profiles is shown by the thick 
solid line which fits well the data (histogram).  
 \label{fig3}}
\end{figure}

The simultaneous fit of a large set of lines with different oscillator
strengths for the same species, and of many different species with 
the same intrinsic line width allows us to constrain the
line broadening parameter (or Doppler width) $b$. 
In other words, a single $b$ value is derived which gives 
the best simultaneous 
fit to the profile of every detected line of each species.
As an example, Fig.~\ref{fig3} shows the fit to the  
O {\sc i} $\lambda$ 1039 line. From the {\sl FUSE} spectrum alone, we obtain
$b=12.0^{+2.0}_{-1.2}$~km~s$^{-1}$. 
The error bar
on the $b$ value includes the uncertainties in the continuum level, the
background residual, the instrumental line spread function and the column
densities of the different species detected at the redshift of
SBS\,0335--052. If the {\sl FUSE} and {\sl HST}
spectra are taken together, we obtain
$b=11.3^{+1.4}_{-1.2}$~km~s$^{-1}$. The two $b$ values are in good agreement 
and thus $b$ is well constrained.

A $b$ value of 12~km~s$^{-1}$ would correspond to a velocity dispersion
$\sigma$ of only $b$/$\sqrt2$ = 8.5~km~s$^{-1}$, somewhat smaller than the
value of $\sim$ 15~km~s$^{-1}$ derived from the H\,{\sc i} VLA map of
\sbs\ \citep{P01}, although the two values are not directly
comparable: the radio observations measure the velocity dispersion within
a region equal to the VLA beam size of 20\farcs5$\times$15\farcs0,
while the {\sl FUSE} observations probe only the star-forming region of $\sim$
2\arcsec\ in size.
Note that if $b$ is underestimated, the column densities derived
from very saturated lines would be overestimated. For instance, we
estimate that if the real value of the $b$ parameter is 16\,km\,s$^{-1}$,
the column density of O\,{\sc i} would be overestimated by $\sim$1~dex. 
As for C\,{\sc ii}, the species for which the column density estimate is
most sensitive to the $b$ value, it would be overestimated by $\sim$1.3~dex. 
Other species are less sensitive to the
exact value of $b$, making the determination of their column
densities more robust.

\subsection{Modeling}

We use the CLOUDY code \citep[ version c90.05]{F96,F98}
to construct a photoionized H {\sc ii} region model that best reproduces the
optical nebular emission-line intensities observed in \sbs\
\citep{I97}. By comparing the H {\sc ii}  column densities predicted by CLOUDY
with the (H {\sc i}  + H {\sc ii}) column densities observed by {\sl FUSE},
we will be able
to deduce the relative amounts of heavy elements in the H {\sc i}  and
H {\sc ii} gas.

We consider a spherically symmetric ionization-bounded H {\sc ii} region
model. The calculations
are stopped in the zone away from the ionizing stars
where the temperature drops to 2000 K.
The ionization in this zone is very low and it is taken to be the outer
edge of the H {\sc ii} region.
Several input parameters need to be set. For a distance of 54.3 Mpc and using
the aperture-corrected H$\beta$ flux from \citet{I97},
we set the H$\beta$ luminosity to be $L$(H$\beta$) = 7.9 $\times$ 10$^{40}$
erg s$^{-1}$, and the number of ionizing photons to be
$N$(Lyc) = 1.7 $\times$
10$^{53}$ s$^{-1}$.
We use the \citet{K91} stellar atmosphere models and adopt the
effective temperature of the ionizing stellar radiation to be
 $T_{\rm eff}$ = 50000 K,
a typical value for low-metallicity high-excitation H {\sc ii} regions.
For the
inner radius of the H {\sc ii} region, we adopt $R_{\rm in}$ =
1.0 $\times$ 10$^{19}$ cm.
The chemical composition of the H {\sc ii} region is set by the observed
element
abundances derived from optical spectroscopy of \sbs\ \citep{I97},
except for the carbon, silicon and phosphorus abundances. 
For carbon we adopt log C/O =
--0.83, and for silicon log Si/O = --1.60 \citep{IT99}. As for phosphorus, we adopt the solar 
log P/O = --3.42 of \citet{GN96}.
The adopted abundances are shown in Table \ref{Tab2}.

We run the CLOUDY code varying the filling factor $f$ and the electron number
density $N_{\rm e}$ 
in order to obtain the best agreement between the predicted and
observed [O {\sc ii}] and [O {\sc iii}] emission lines.
We find that if $N_{\rm e}$ varies in the ranges 7 -- 13 cm$^{-3}$ and $f$ 
in the range 0.17 -- 0.29, then we have adequate agreement between 
the observed and model line intensities. 
The best model is found for a filling
factor $f$ = 0.22 and an electron number density $N_{\rm e}$ = 10 cm$^{-3}$.
The optical line intensities of the best model are shown in Table \ref{Tab3}, 
where
we compare them with the observed ones \citep{I97}. The error bars of
the model line intensities (and those of the model parameters in Tables \ref{Tab1} 
and \ref{Tab4}) correspond to the maximum range of their values.
There is good general agreement between observed and model 
line intensities, giving us confidence that the
photoionization model is correct.
The predicted total hydrogen (neutral and ionized) column density in the
 H {\sc ii} region is
$N$(H {\sc i} + H {\sc ii}) = 4.93 $\times$ 10$^{21}$ cm$^{-2}$ or
log $N$(H {\sc i} + H {\sc ii})= 21.69.
As expected for a H {\sc ii} region model, most of the hydrogen is ionized:
 $N$(H {\sc ii}) = 4.89 $\times$ 10$^{21}$ cm$^{-2}$. The column density
of the neutral hydrogen is two orders of magnitude lower. The model also
predicts a column density of molecular hydrogen made via H$^-$
$N$(H$_2$) = 1.91 $\times$ 10$^{11}$ cm$^{-2}$, more than six orders of
magnitude lower than our observational upper limit.

\begin{figure}
\epsscale{0.8}
\plotone{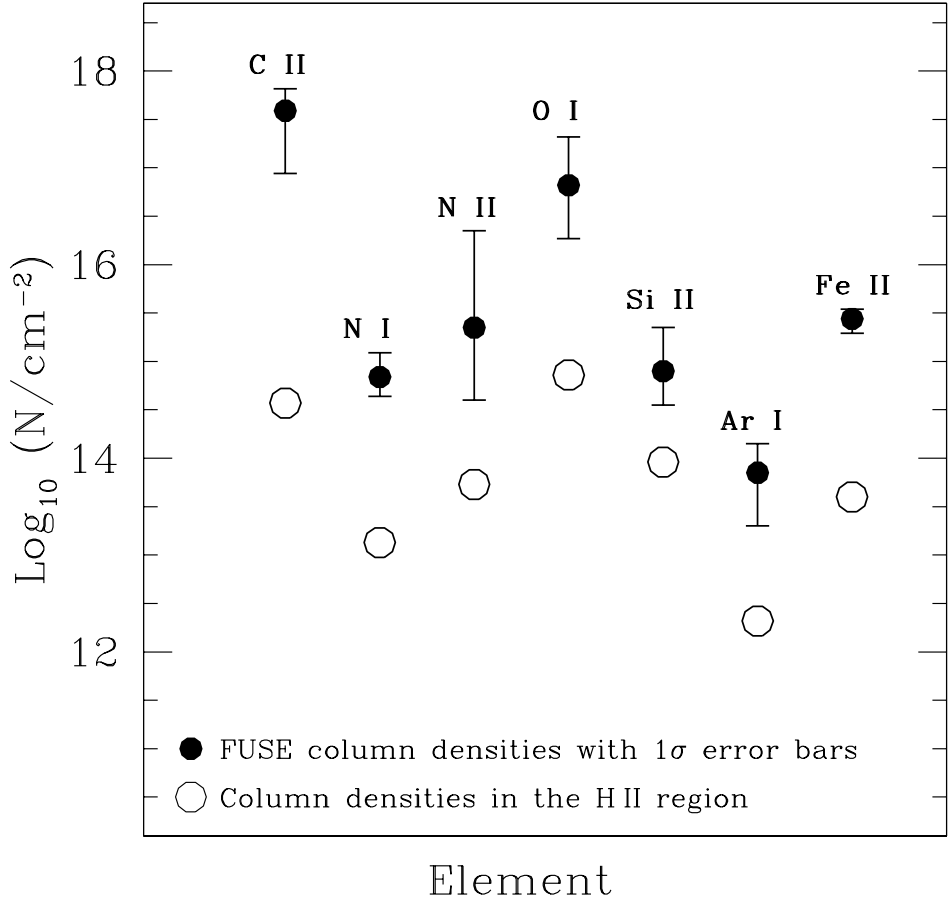}
\figcaption{Heavy element column densities in the H {\sc i} and H {\sc ii} gas 
of \sbs.
The filled circles show the column densities derived from the {\sl FUSE}
data. The open circles show the column densities in the H\,{\sc ii}
region as calculated from CLOUDY modeling of the optical emission
lines. The column densities in the H\,{\sc i} cloud
follow the same pattern as the ones in the H {\sc ii} gas, although they are 
larger by 1-2 orders of magnitude. 
While this pattern reflects the cosmic abundances of the considered
species, the fact that we recover this pattern shows
the consistency of the measured column densities in the H\,{\sc i} gas 
although they have been derived from low S/N data. 
The lower column densities in the ionized 
gas show that the absorption lines in the {\sl FUSE} spectra arise  
in the neutral interstellar gas.
\label{fig4}}
\end{figure}

We now compare the column densities of heavy elements predicted by
CLOUDY with those derived from the {\sl FUSE} and {\sl HST} spectra.
The CLOUDY predicted column densities are given in Table \ref{Tab4}. Those relevant
to the {\sl FUSE} observations are also repeated in column 3 of Table \ref{Tab1}.
In Table \ref{Tab4}, $x$ is the radially averaged ratio of the ion number to the
total number of a particular element, e.g., $N$(Fe$^+$)/$N$(Fe). The
predicted log $N$(X) of species X is then derived as log $N$(X) = log
$N$(H~{\sc i}+H {\sc ii}) + log X$^{+i}$/H -- log $x$(X$^{+i}$). The
column densities
listed in Table \ref{Tab1} are plotted with their 1 $\sigma$ error bars as filled
circles in Fig. \ref{fig4}. For comparison, the corresponding CLOUDY predicted
column densities are plotted as open circles. There are two features that
should be noted. First, the observed and calculated relative variations
from ionic element to ionic element show the same pattern. While this 
simply reflects the relative cosmic abundances of the considered 
species, retrieving that pattern from the data does show the consistency 
of the derived column densities for different species despite the
low signal-to-noise ratio of our {\sl FUSE} spectra.

Second, the majority of
the calculated H {\sc ii} column densities are systematically smaller by
more than one order of magnitude than the observed column densities. 
This implies
that the {\sl FUSE} column densities arise mainly in the H {\sc i} neutral
envelope, and not in the gas ionized by young stars.

\begin{figure}
\epsscale{0.8}
\plotone{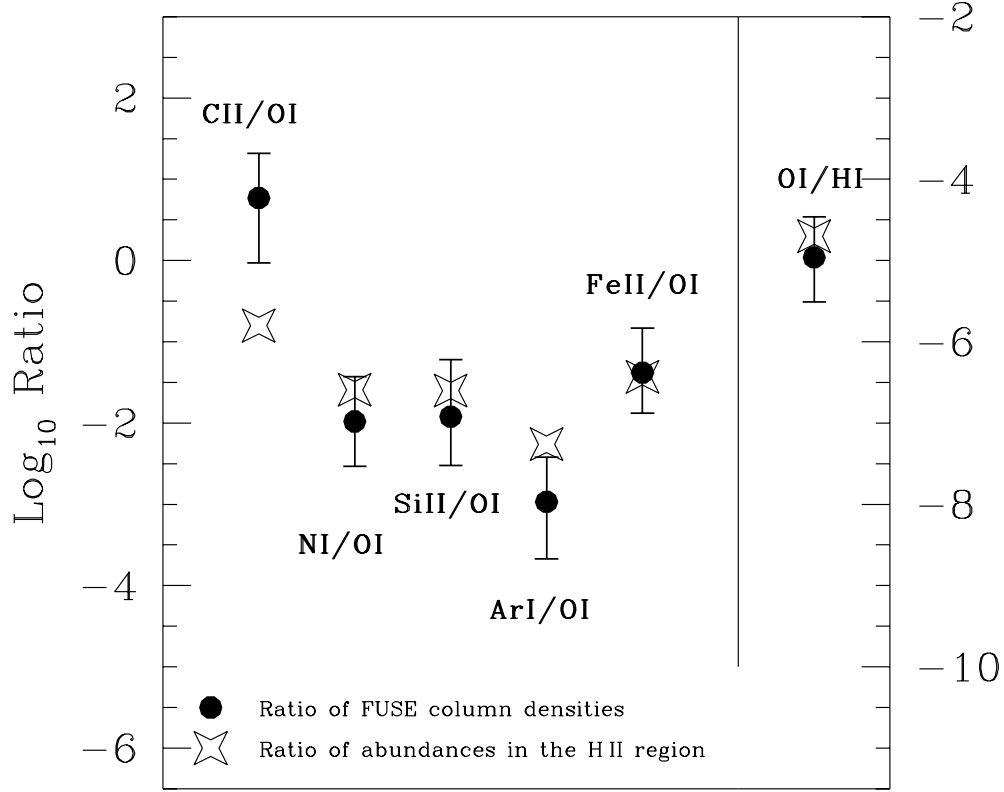}
\figcaption{
Relative ratios of {\sl FUSE} column densities (filled circles) in the 
H {\sc i} gas 
compared to the corresponding abundance ratios in the H {\sc ii} gas (stars)
in \sbs. 
For each {\sl FUSE} column density, the ionization state shown is 
the main one in the H {\sc i} region and the 
ionization correction factors are very near unity. For the  H {\sc ii}
region, total element abundances
including all ionization states are plotted. From left to right 
are the C/O, N/O, Si/O, Ar/O, Fe/O and O/H ratios. The right ordinate
is for the O/H ratio, while the left ordinate is for all other abundance
ratios.
Note in particular that
the O/H ratios in the H\,{\sc i} and
H\,{\sc ii} regions are very similar. \label{fig5}}
\end{figure}

Fig. \ref{fig5} shows the ratio of the column
densities of all ions relative to that of O {\sc i}. As for the
column density of O {\sc i}, it is given relative to that of H {\sc i}.
There is general good agreement between the ratios of the {\sl FUSE}
column densities (filled circles) and the ratios of the elements abundances
in the H {\sc ii} region (stars). The only exception is the 
$N$(C {\sc ii})/$N$(O {\sc i}) ratio which is a factor of 30 larger than the 
C/O ratio in
the H {\sc ii} region. As discussed in Section 5.1, the C {\sc ii} column density 
is derived from a single strongly saturated line and is probably 
overestimated.  

Of special interest is the $N$(O {\sc i})/$N$(H {\sc i}) ratio. 
The ionization potential
of O\,{\sc i} is very similar to that of H\,{\sc i}, and there is an efficient
charge exchange between O\,{\sc ii} and H\,{\sc i}. 
Hence there is a strong coupling between the oxygen
and hydrogen ionization fractions.
In the neutral gas, the O\,{\sc i}/H\,{\sc i} ratio can therefore
be considered as a very good proxy for the O/H ratio and hence it is a 
very good tracer of 
metallicity. We obtain log $N$(O {\sc i})/$N$(H {\sc i}) =
 --5.04$\pm$0.55
or [O {\sc i}/H {\sc i}] = --1.70. 
This is to be compared with log O/H = --4.70$\pm$0.01 for the ionized gas 
in \sbs\ \citep{I99}. Thus, within the errors, the
metallicity of the neutral gas in SBS 0335--052 is 
comparable to that of its ionized gas. The much lower
metallicity of the neutral gas of SBS 0335--052 
obtained by \citet{TI97} from {\sl HST} observations is not correct as it is 
based on the erroneous assumption that the O {\sc i}, Si {\sc ii} and 
S {\sc ii} lines are not saturated.   

How does the
metallicity of the H {\sc i} gas in SBS 0335--052 compare with those
determined in the neutral ISM of other BCDs? Such a determination has been
made for 
four other BCDs. For the most metal-deficient BCD known, I Zw 18
with a log (O/H)$_i$ of the ionized gas equal to --4.82$\pm$0.01 
\citep{IT99}, 
\citet{A03} have obtained 
log (O/H)$_n$ of the neutral gas = --5.37$\pm$0.28
while \citet{L04} derived the higher value log (O/H)$_n$ =
--4.7$\pm$0.35. 
The discrepancy between the two values may be
due to the fact that, to derive O/H, \citet{A03} use 
the strongly saturated
O\,{\sc i} line at $\lambda_0=1039$ \AA\ which, at the redshift of I Zw 18,
is blended with terrestrial airglow. 
For I~Zw~36 $\equiv$ Mrk 209 with log (O/H)$_i$ = --4.23$\pm$0.01 \citep{IT99},
\citet{Lebouteiller} derived
log (O/H)$_n$ = --4.5$^{+1.0}_{-0.6}$.
%
For the BCD Mrk 59 with log (O/H)$_i$ = --4.01$\pm$0.01 
\citep{IT99}, \citet{T02} have derived log (O/H)$_n$ = --5.0$\pm$0.3.
As for the BCD NGC 1705 with log (O/H)$_i$ = --3.79$\pm$0.05 
\citep{LS04}, \citet{H01} have derived log (O/H)$_n$ = --4.6$\pm$0.3. 
Thus if we disregard the low value of \citet{A03}, 
the ionized gas metallicity of 
the five different BCDs observed by {\sl FUSE} 
span a log (O/H)$_i$ range from --4.82 to --3.79, i.e. a 1.03 dex spread. 
On the other hand,  
the metal content in their H {\sc i} envelopes
varies in the range 
log (O/H)$_n$ = --5.0 - --4.5,  
or [O/H]$_n$ = --1.7 - --1.2, using the revised solar abundance (O/H)$_\odot$ = 
--3.34 of \citet{A04}, i.e. a 0.5 dex spread. 

At first glance, the 
metallicity spread of the neutral gas component appears to be 
only slightly lower than that of the ionized gas component.  
However a real difference appears once the error bars 
of the metallicity measurements are considered. 
The metallicities of the ionized gas, log (O/H)$_i$, are 
determined from emission-line spectra 
with a precision of 0.01 dex (0.05 dex for NGC 1705), 
while those of the neutral gas, 
log (O/H)$_n$, derived from absorption spectra, 
have considerably larger error bars varying from 0.3 dex 
for Mrk 59 and NGC 1705 to 1.0 dex for I Zw 36. 
These large error bars are in fact  
responsible for the apparent large metallicity spread of the neutral gas.      If we take into account the error bars, the log (O/H)$_i$ values do vary 
over a 1.0 dex range while the log (O/H)$_n$ measurements
are consistent with a single constant value.
The spread of the log (O/H)$_n$ measurements around their mean value
$<\log ({\rm O/H})_n> = -4.8$ can be evaluated by the sum of their 
squared normalized
differences which, for five measurements, follow a $\chi^2_4$ law with four
degrees of freedom. We obtain a very low dispersion
$\sum \left[ (\log ({\rm O/H})_n - <\log ({\rm O/H})_n>)/\sigma \right]^2 = 
1.40$. There is more than 84\% chance that  
$\chi^2_4$ is larger than this value,
supporting our contention that all five measurements are consistent with 
a single value of log (O/H)$_n$. 

The fact that the most metal-deficient BCD (I Zw 18) has 
about the same log (O/H)$_n$ as the most metal-rich BCD (NGC 1705) studied 
thus far by {\sl FUSE} 
suggests that the metallicities of the neutral
and ionized gas are unrelated. 
As discussed in the Introduction, the two most metal-deficient BCDs known, 
I Zw 18 and SBS 0335--052 have been suggested to be bona fide young
galaxies, with an age not exceeding 500 Myr 
[see \citet{IT04} for
a discussion of the age of I Zw 18 and TIL97, \citet{P98} and \citet{P04} for
a discussion of the age of SBS 0335--052]. Thus the H {\sc i} envelopes
that surround these two BCDs may be expected to be truly primordial, i.e.
not to contain heavy elements. Our study here of \sbs\
 and previous work on I Zw 18 show
that this is clearly not the case, that the H {\sc i} envelopes in these
two BCDs have been enriched previously to a level 
[O/H]$_n$ = --1.7 - --1.4,
perhaps by Population III stars. This speculation appears to be supported
by the oxygen abundances derived by \citet{Te02} for the
intergalactic medium using ultraviolet absorption lines in Ly$\alpha$
absorbers. Those authors derive a [O/H] range of --2.0 to --1.1 [we have 
corrected their values using the O solar abundance of \citet{A04}], 
which
overlaps well with that derived thus far for the neutral ISM in five BCDs.


        \section{Conclusions}

We present {\sl FUSE} far-UV spectra 
of the second most metal-deficient 
Blue Compact Dwarf (BCD) galaxy known, SBS 0335--052 with an 
ionized gas oxygen abundance log O/H = --4.70$\pm$0.01.
Because this galaxy is one of the best candidates
for being a young galaxy in the local universe, with an age not exceeding 
$\sim$ 100 Myr, the study of the abundances in its neutral hydrogen 
envelope is of special interest as it can shed light on the possible  
metal enrichment of the intergalactic medium by Population III stars several
hundred million years after the Big Bang. We have obtained the following 
results:   

1. No H$_2$ absorption lines are detected. We set 
a 3$\sigma$ upper limit for diffuse H$_2$ in \sbs\ of 
3.7$\times$10$^{17}$ cm$^{-2}$ for $T$ = 500 K and 5.5$\times$10$^{17}$ 
cm$^{-2}$ for $T$ = 1000 K. The absence of diffuse H$_2$ is due to the 
combined effects of a low H\,{\sc i} density, a 
large UV flux which destroys H$_2$ molecules  
and a low metallicity which makes grains on which to form H$_2$ molecules 
scarce. 

2. The non-detection of diffuse H$_2$ contrasts with the detection of H$_2$ 
emission lines in near-infrared spectra of \sbs.
This implies that the detected H$_2$ must be clumpy, and associated only 
with the dense star-forming regions. These clumps should be denser than 
$\sim$ 1000 cm$^{-3}$ and have a temperature greater than $\sim$ 1000 K.
For such dense clumps, H$_2$~self-shielding 
can protect the H$_2$ molecules from dissociating UV-photons.
Each clump has a mass less than $\sim$ 5$\times$10$^4$ $M_\odot$. There 
should be at least $\sim$ 2000 clumps,  
accounting for $\geq$ 5\% of the total H {\sc i} mass.  

3. 
The O\,{\sc i}/H\,{\sc i} ratio in the H {\sc i} gas 
(log $N$(O {\sc i})/$N$ (H {\sc i}) =
--5.04$\pm$0.55 is similar, within the errors,
 to the O/H ratio derived from optical emission lines in 
the H\,{\sc ii} region of \sbs\ (log O/H = --4.70$\pm$0.01), i.e.  
the metallicity of the neutral gas is comparable to that of the 
ionized gas. Thus although \sbs\ is a probable young galaxy, its neutral 
envelope is not pristine.

4. There exists a noteworthy similarity between the O/H
ratio observed in the H {\sc i} gas 
of SBS\,0335--052 (with log (O/H)$_i$ of the ionized gas equal to --4.70) 
and the ones seen in 
the H {\sc i} gas of 
other BCDs for which similar {\sl FUSE} studies 
have been carried out, I Zw 18 [log (O/H)$_i$ = --4.82], I Zw 36 
[log (O/H)$_i$ = --4.23],
Mrk\,59 [log (O/H)$_i$ = --4.01] and NGC 1705 [log (O/H)$_i$ = --3.79]. 
In all five BCDs, log O/H of the neutral gas is in the range  
--5.0 - --4.5 (or [O/H] = --1.7 - --1.2), 
although the ionized gas of 
NGC 1705 is some 10 times more metal-rich than that of \sbs\ and I Zw 18. 
 This suggests that the metallicities of the neutral and ionized gas 
components are unrelated.
We may speculate that the neutral gas in these five  
BCDs has been enriched to a common metallicity level of log (O/H) $\sim$ --5 
by population~III~stars. 
This speculation appears to be 
supported by observations of Ly$\alpha$ absorbers which give  
similar oxygen abundances in the intergalactic medium.

\acknowledgements
T.X.T. thanks the hospitality of the Institut d'Astrophysique de Paris. 
A.L. and Y.I.I. acknowledge the hospitality of the Astronomy Department of the 
University of Virginia.
T.X.T. has been supported in part by NASA grant NAG5-10341.
T.X.T and Y.I.I.
are grateful for the partial financial support of the U.S. Civilian
Research \& Development Foundation for the Independent States of the
Former Soviet Union (CRDF) through Award No. UP1-2551-KV-03, and of the
National Science Foundation through grant AST-02-05785.
This work has been done using the profile fitting procedure {\tt Owens.f}
developed by M.~Lemoine and the {\sl FUSE} French Team.



\clearpage

%

\begin{deluxetable}{llc}
\tablenum{1}
\tablecolumns{2}
\tablewidth{0pt}
\tablecaption{Heavy element column densities (cm$^{-2}$) in SBS 0335--052
\label{Tab1}}
\tablehead{
Species   & log\,$N$ (cm$^{-2}$)\tablenotemark{a} &  CLOUDY\tablenotemark{b}} 
\startdata
H {\sc i} ({\sl FUSE})      &21.86$^{+0.08}_{-0.05}$&21.69$^{+0.12}_{-0.11}$
\\
C {\sc ii}                  &17.59$^{+0.23}_{-0.65}$ &14.55$^{+0.02}_{-0.02}$ \\
N {\sc i}                   &14.84$^{+0.25}_{-0.2}$  &13.13$^{+0.01}_{-0.00}$ \\
N {\sc ii}                  &15.35$^{+1.0}_{-0.75}$  &13.73$^{+0.02}_{-0.01}$ \\
N {\sc ii}$^*$              &13.70$^{+0.25}_{-13.70}$&      \\
N {\sc iii}                 &18.70$^{+1.0}_{-18.7}$ &15.29$^{+0.10}_{-0.10}$ \\
O {\sc i}                   &16.82$^{+0.5}_{-0.6}$  &14.86$^{+0.01}_{-0.01}$ \\
Si {\sc ii}                 &14.90$^{+0.45}_{-0.35}$  &13.96$^{+0.04}_{-0.04}$ \\
Ar {\sc i}                  &13.85$^{+0.3}_{-0.55}$  &12.32$^{+0.01}_{-0.01}$ \\
P {\sc ii}                  &14.00$^{+1.0}_{-1.5}$    &12.06$^{+0.01}_{-0.01}$ \\
P {\sc ii}$^*$              &13.09$^{+0.5}_{-13.}$  &      \\
Fe {\sc ii}                 &15.44$^{+0.1}_{-0.15}$  &13.60$^{+0.02}_{-0.01}$ \\
\enddata
\tablenotetext{a}{Fitting with a single Voigt profile.
Error bars are 1 $\sigma$ limits. 
}
\tablenotetext{b}{Total H {\sc i} + H {\sc ii} column density in the
H {\sc ii} region.}
\end{deluxetable}

\clearpage

%
\begin{deluxetable}{lc}
\tablenum{2}
\tablecolumns{2}
\tablewidth{0pt}
\tablecaption{Abundances in SBS 0335--052 used as input to
CLOUDY \label{Tab2}}
\tablehead{Species   & log ($N$(species)/$N$(H)) }
\startdata
He                  & ~~~$-1.0915$\tablenotemark{a}  \\
C                   & $-5.50$\tablenotemark{b}  \\
N                   & $-6.26$\tablenotemark{a}  \\
O                   & $-4.67$\tablenotemark{a}  \\
Ne                  & $-5.48$\tablenotemark{a}  \\
Si                  & $-6.27$\tablenotemark{b}  \\
P                   & $-8.05$\tablenotemark{c}  \\
S                   & $-6.23$\tablenotemark{a}  \\
Ar                  & $-6.93$\tablenotemark{a}  \\
Fe                  & $-6.10$\tablenotemark{a}  \\
\enddata
\tablenotetext{a}{From \citet{I97}.}
\tablenotetext{b}{From \citet{IT99}.}
\tablenotetext{c}{Adopting solar log P/O = --3.42 from \citet{GN96}.}
\end{deluxetable}

\clearpage

%
\begin{deluxetable}{lcc}
\tablenum{3}
\tablecolumns{3}
\tablewidth{0pt}
\tablecaption{Comparison between the observed and CLOUDY predicted
optical line intensities normalized to H$\beta$ \label{Tab3}}
\tablehead{ Ion  &  Observed\tablenotemark{a} & CLOUDY }
\startdata
3727 [O {\sc ii}]       &      0.233  & 0.217$^{+0.046}_{-0.041}$ \\

3835 H9                 &      0.079  & 0.079$^{+0.000}_{-0.000}$ \\
3868 [Ne {\sc iii}]     &      0.239  & 0.253$^{+0.014}_{-0.014}$ \\
3889 He {\sc i} + H8    &      0.172  & 0.194$^{+0.002}_{-0.001}$ \\
3968 [Ne {\sc iii}] + H7&      0.239  & 0.239$^{+0.004}_{-0.003}$ \\
4101 H$\delta$          &      0.255  & 0.264$^{+0.000}_{-0.000}$ \\
4340 H$\gamma$          &      0.476  & 0.474$^{+0.001}_{-0.000}$ \\
4363 [O {\sc iii}]      &      0.109  & 0.102$^{+0.008}_{-0.009}$ \\
4471 He {\sc i}         &      0.035  & 0.035$^{+0.001}_{-0.000}$ \\
4658 [Fe {\sc iii}]     &      0.003  & 0.002$^{+0.001}_{-0.000}$ \\
4740 [Ar {\sc iv}]      &      0.009  & 0.007$^{+0.001}_{-0.001}$ \\
4861 H$\beta$           &      1.000  & 1.000$^{+0.000}_{-0.000}$ \\
4959 [O {\sc iii}]      &      1.054  & 1.099$^{+0.061}_{-0.060}$ \\
5007 [O {\sc iii}]      &      3.155  & 3.173$^{+0.172}_{-0.175}$ \\
5876 He {\sc i}         &      0.100  & 0.095$^{+0.000}_{-0.000}$ \\
6300 [O {\sc i}]        &      0.007  & 0.002$^{+0.000}_{-0.001}$ \\
6312 [S {\sc iii}]      &      0.006  & 0.013$^{+0.000}_{-0.001}$ \\
6563 H$\alpha$          &      2.745  & 2.840$^{+0.007}_{-0.008}$ \\
6584 [N {\sc ii}]       &      0.007  & 0.004$^{+0.001}_{-0.000}$ \\
6678 He {\sc i}         &      0.027  & 0.027$^{+0.000}_{-0.000}$ \\
6717 [S {\sc ii}]       &      0.019  & 0.013$^{+0.002}_{-0.003}$ \\
6731 [S {\sc ii}]       &      0.017  & 0.009$^{+0.002}_{-0.001}$ \\
7065 He {\sc i}         &      0.039  & 0.024$^{+0.001}_{-0.001}$ \\
7136 [Ar {\sc iii}]     &      0.014  & 0.029$^{+0.001}_{-0.001}$ \\
\enddata
\tablenotetext{a}{From \citet{I97}.}
\end{deluxetable}

\clearpage

%
\begin{deluxetable}{lcc}
\tablenum{4}
\tablecolumns{3}
\tablewidth{0pt}
\tablecaption{CLOUDY predicted column densities in the H {\sc ii} region \label{Tab4}}
\tablehead{Ion   & log $x$\tablenotemark{a} & log $N$\tablenotemark{b} }
\startdata
C {\sc ii}~~~~~~~~~~~~~~~~~~~~~~~~~~~~~~   &   $-1.640^{+0.091}_{-0.093}$ & 14.55$^{+0.02}_{-0.02}$ \\
N {\sc i}    &   $-2.295^{+0.108}_{-0.114}$ & 13.13$^{+0.01}_{-0.00}$ \\
N {\sc ii}   &   $-1.700^{+0.096}_{-0.102}$ & 13.73$^{+0.02}_{-0.01}$ \\
N {\sc iii}  &   $-0.141^{+0.011}_{-0.014}$ & 15.29$^{+0.10}_{-0.10}$ \\
O {\sc i}    &   $-2.164^{+0.100}_{-0.106}$ & 14.86$^{+0.01}_{-0.01}$ \\
Si {\sc ii}  &   $-1.458^{+0.073}_{-0.077}$ & 13.96$^{+0.04}_{-0.04}$ \\
P {\sc ii}   &   $-1.584^{+0.102}_{-0.109}$ & 12.06$^{+0.01}_{-0.01}$ \\
Ar {\sc i}   &   $-2.436^{+0.104}_{-0.110}$ & 12.32$^{+0.01}_{-0.01}$ \\
Fe {\sc ii}  &   $-1.991^{+0.098}_{-0.103}$ & 13.60$^{+0.02}_{-0.01}$ \\
\enddata
\tablenotetext{a}{Radially averaged ratio of the ion number to the total
number of the element.}
\tablenotetext{b}{Column density in cm$^{-2}$.}
\end{deluxetable}


\clearpage


\clearpage


\clearpage


\clearpage


\clearpage

\end{document}